\documentclass[aip,jap,reprint]{revtex4-1}
\usepackage[dvips]{graphicx}
\usepackage{amsmath}

\begin{document}

\title{Charge carrier dynamics in bulk MoS$_2$ crystal studied by transient absorption microscopy}

\author{Nardeep Kumar}
\affiliation{Department of Physics and Astronomy, The University of Kansas, Lawrence, Kansas 66045, USA}

\author{Jiaqi He}
\affiliation{Key Laboratory of Luminescence and Optical Information, Ministry of Education, Institute of Optoelectronic Technology, Beijing Jiaotong University, Beijing 100044, China}

\author{Dawei He}
\affiliation{Key Laboratory of Luminescence and Optical Information, Ministry of Education, Institute of Optoelectronic Technology, Beijing Jiaotong University, Beijing 100044, China}

\author{Yongsheng Wang}\email{yshwang@bjtu.edu.cn}
\affiliation{Key Laboratory of Luminescence and Optical Information, Ministry of Education, Institute of Optoelectronic Technology, Beijing Jiaotong University, Beijing 100044, China}

\author{Hui Zhao}\email{huizhao@ku.edu}
\affiliation{Department of Physics and Astronomy, The University of Kansas, Lawrence, Kansas 66045, USA}

\date{\today}

\begin{abstract}
We report a transient absorption microscopy study of charge carrier dynamics in bulk MoS$_2$ crystals at room temperature. Charge carriers are injected by interband absorption of a 555-nm pulse, and probed by measuring differential reflection of a time-delayed and spatially scanned 660-nm pulse. We find an intervalley transfer time of about 0.35~ps, an energy relaxation time of hot carriers on the order of 50 ps, and a carrier lifetime of 180~$\pm$~20~ps. By monitoring the spatiotemporal dynamics of carriers, we obtained a diffusion coefficient of thermalized electrons of 4.2 $\pm$ 0.5 cm$^2$/s, corresponding to a mobility of 170 $\pm$ 20~cm$^2$/Vs. We also observed a time-varying diffusion coefficient of hot carriers.
\end{abstract}

\maketitle

\section{Introduction}

Layered materials in which atomic sheets are stacked together by weak van der Waals forces can be used to fabricate two-dimensional systems. They represent a diverse and rich, but largely unexplored, source of materials. Graphene is the most well-known example.\cite{pnas10210451,rpp74082501,s306666,s3121191,n457706,nn4712} However, many other layered materials can also be explored. Very recently, significant progress has been made on one type of layered transition metal dichalcogenide, MoS$_2$. In 2010, two groups independently discovered an indirect-to-direct bandgap transition that occurs when varying the thickness from bulk to monolayer.\cite{l105136805,nl101271} In 2011, monolayer MoS$_2$ transistors with a 10$^8$ on/off ratio and a room-temperature mobility of more than 200~cm$^2$/Vs have been demonstrated.\cite{nn6147} In 2012, several groups reported observation of valley selective optical excitation and luminescence in monolayer MoS$_2$.\cite{nc3887,nnano7490,nnano7494,l108196802,b86081301} These studies paved ways to apply two dimensional MoS$_2$ for photonics, electronics, and valleytronics. 

Studies on {\it bulk} MoS$_2$ can provide complementary information for understanding monolayers and their interaction with environments and substrates, since the properties of monolayer MoS$_2$ are different from, but related to, the bulk. Here we report a transient absorption microscopy study of charge carrier dynamics in bulk MoS$_2$ crystals. By time resolving the dynamics, we observed intervalley transfer, energy relaxation, and recombination of carriers. Furthermore, spatial resolution of carrier dynamics allows us to observe diffusion of carriers, and measure their diffusion coefficients. These results provide fundamental information on charge carrier dynamics in bulk MoS$_2$ crystals, and can be used to understand carrier dynamics of MoS$_2$ bulk and atomic layers.

\section{Experimental scheme and expected carrier dynamics}

\begin{figure}
 \includegraphics[width=8.5cm]{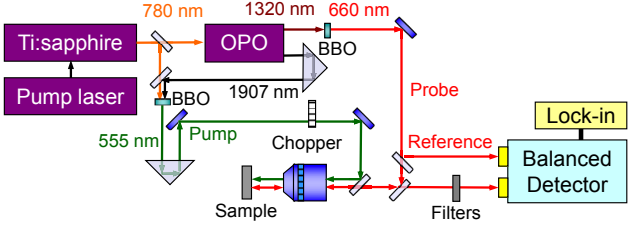}
 \caption{Schematics of the transient absorption microscopy system.}
  \label{setup}
\end{figure}

Figure~\ref{setup} shows schematically the transient absorption microscopy setup. A passive mode-locked Ti:sapphire laser is used to generate 100-fs pulses with a central wavelength of 780 nm and a repetition rate of 80 MHz. The output is used to pump an optical parametric oscillator (OPO), which has a signal output of 1320 nm and an idler output of 1907 nm. We take a portion of the 780-nm Ti:sapphire output before entering the OPO, and combine it with the idler beam of OPO. By sending both beams to a beta barium borate (BBO) crystal, we generate their sum frequency with a central wavelength of 555-nm. This pulse is focused to the sample surface through a microscope objective lens with a spot size of 1.6 $\mu$m (in full width at half maximum). It is used as the pump pulse. The probe pulse is obtained by second harmonic generation of the 1320 nm signal output of OPO in another BBO crystal, with a central wavelength of 660 nm. It is focused to the sample through the same objective lens, with a spot size of 1.3 $\mu$m. The reflected probe is collimated by the objective lens, and is sent to one photodiode of a balanced detector. A portion of the probe beam is taken before entering the sample, and is sent to the other photodiode of the balanced detector as the reference beam. 

The balanced detector outputs a voltage that is proportional to the difference between optical powers on the two photodiodes. In the measurements, we first block the pump pulse, and adjust the reference to match the reflected probe power, such that the balanced detector outputs a zero voltage. We then allow the pump pulse to reach the sample. It injects carriers, which change the reflection coefficient of the sample for the probe pulse. Hence, the balanced detector now outputs a voltage that is proportional to a differential reflection of the probe, $\Delta R/R_0$. It is defined as the relative change of the probe reflection caused by the pump, $(R-R_0)/R_0$, where $R$ and $R_0$ are the reflection of the probe with the pump presence and without it, respectively. The advantage of using the balanced detector is to suppress the intensity noise of the probe pulse. Such a common mode noise is equally distributed on the two photodiodes, and hence is almost cancelled. To suppress other types of noise, we use a mechanical chopper to modulate the pump intensity at about 2 KHz, and use a lock-in amplifier to read the output of the balanced detector at that frequency. To measure the differential reflection as a function of the probe delay, defined as the time delay of the probe pules with respect to the pump pulse, we change the length of the pump arm by moving the retroreflector in the pump arm. The time resolution in these measurements is about 250 fs, since both pulses are about 180 fs at sample, mainly due to dispersion of the objective lens. To spatially resolve the differential reflection signal, we scan the pump spot with respect to the probe spot at the sample surface by tilting the beamsplitter that reflects the pump into the objective lens.

The MoS$_2$ samples used for this study are natural occurring crystals from SPI Supplies. The background carrier density is below $10^{15}$ /cm$^3$. The samples were mounted on glass substrates. The surface of the samples is mechanically cleaved by using an adhesive tape in order to obtain a relative flat, clean, and fresh surface. All the measurements were performed under ambient conditions. 

\begin{figure}
 \includegraphics[width=8.5cm]{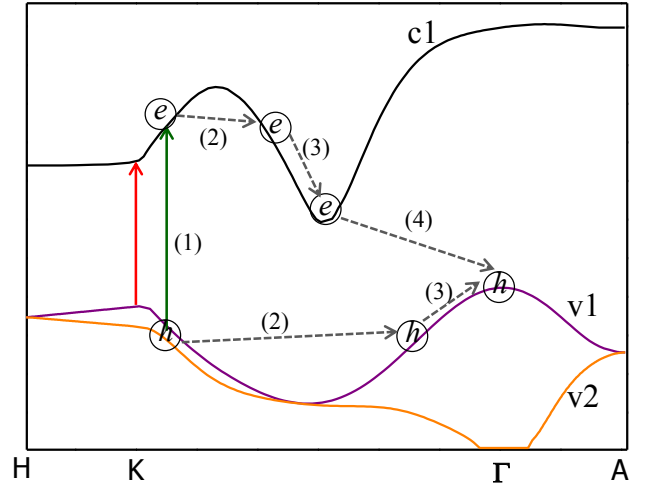}
 \caption{Bandstructure of MoS$_2$ including the conduction (C1) and valence (V1 and V2) bands. The vertical green and red arrows illustrate the pump and probe photon energies. The expected carrier dynamics is also shown, including interband excitation (1), intervalley transfer (2), energy relaxation (3), and interband recombination (4).}
  \label{energybands}
\end{figure}

The bandstructure of MoS$_2$ is plotted schematically in Fig.~\ref{energybands}, along with the expected carrier dynamics. The pump photon energy of 2.24 eV is large enough to excite electrons near the direct bandgaps at K point, from both $V$1 and $V$2 valanced bands to the C1 conduction band. Once excited, the carriers can rapidly transfer to the lower valleys by intervalley scattering. These carriers then undergo an energy relaxation process, via emitting phonons, to reach thermal equilibrium with the lattice. Finally, the electrons and holes recombine. The dynamics is detected by using the 660-nm probe, with a photon energy of 1.88 eV, within the A-exciton resonance.\cite{l105136805,nl101271} Although the probe is resonant with the exciton transition, we do not expect it detects real exciton population. Rather, the probe detects free carriers via carrier-induced nonlinearity at exciton transition.\cite{b326601} In bulk MoS$_2$, the exciton binding energy is about 25 meV, which is comparable to thermal energy at room temperature. \cite{b85205302} Hence, we do not expect a significant exciton population, and ignore their contributions. In fact, our experimental results also confirm this assumption (see Sec.~\ref{discussion}). 

\section{Differential reflection to probe carrier dynamics}

In this section, we discuss in detail how to monitor the carrier dynamics with the probe pulse. For the normal incident probe pulse, the linear reflection coefficient of the unexcited sample can be obtained from Snell's law,
\begin{equation}
R_0=\frac{(n_0-1)^2+\kappa_0^2}{(n_0+1)^2+\kappa_0^2},
\end{equation}
where $n_0$ and $\kappa_0$ are the real and imaginary parts of the index of refraction. The latter, known as the extinction coefficient, is related to the absorption coefficient ($\alpha_0$) and light wavelength ($\lambda$) by $\kappa_0 = (\lambda / 4 \pi) \alpha_0$. The charge carriers injected by the pump pulse can change the reflection coefficient to
\begin{equation}
R=\frac{(n-1)^2+\kappa^2}{(n+1)^2+\kappa^2} \equiv R_0 +\Delta R.
\end{equation}
Since in our measurements the probe pulse is tuned to an excitonic resonance, the absorption is expected to play a dominant role. Hence, we assume real part is unchanged, $n=n_0$,  and $\kappa = \kappa_0 + \Delta \kappa$. Furthermore, if $\Delta \kappa$ only causes a small relative change of $R$, {i.e.} $\Delta R \ll R_0$, we have
\begin{equation}
\Delta R = - \frac{8 n_0 \kappa_0}{[(n_0 + 1)^2 + \kappa_0^2]^2} \Delta \kappa. 
\end{equation}
In our measurements, $\Delta R / R_0$ is on the order of $10^{-3}$. Hence, this condition is safely satisfied, and the differential reflection 
\begin{equation}
\frac{\Delta R}{R_0} = - \frac{8 n_0 \kappa_0}{[(n_0 + 1)^2 + \kappa_0^2][(n_0 - 1)^2 + \kappa_0^2]} \Delta \kappa. 
\end{equation}

In our scheme, we probe the injected carriers by saturation of excitonic absorption induced by these carriers. In most cases, excitonic nonlinearities, such as absorption saturation, are induced by real exciton populations via phase-space state filling. However, it has been established that excitonic transition strength can be changed by the phase-space state filling effect of free carriers, since the exciton wave function is composed of free-carrier states.\cite{b326601} Although this effects is generally weaker than those caused by exciton population, it is less selective to carrier energy distribution. Hence, it can effectively probe carriers with a high kinetic energy. Recently, we have shown that free carrier induced excitonic nonlinearity can be used to probe ballistic transport of hot carriers\cite{l106107205,b79155204,b79115321,b78045314,l96246601,b75075305} and coherent plasma oscillation\cite{jap103053510} in GaAs.  In this effect, the change in extinction coefficient can be described by
\begin{equation}
\Delta \kappa = - \sigma \kappa_0 \frac{N}{N+N_s},
\end{equation}
where $N$ is the carrier density, $N_s$ is the saturation density and the cross section $\sigma$ is introduced to describe the effectiveness of carriers in saturating the probe transitions. Therefore,  the differential reflection is related to the carrier density by
\begin{equation}
\frac{\Delta R}{R_0} =  \frac{8 n_0 \kappa_0^2}{[(n_0 + 1)^2 + \kappa_0^2][(n_0 - 1)^2 + \kappa_0^2]} \sigma \frac{N}{N+N_s}. 
\label{DRoR}
\end{equation}

In the experiments, carriers are injected by interband absorption of the pump pulse. The on-axis peak energy fluence of the pump pulse, $F_0$, is related to its time average power, $P$, by
\begin{equation}
F_0 = \frac{4 \mathrm{ln}(2) P}{\pi f w^2},
\end{equation}
where $f$ is the repetition rate of the laser and $w$ is the full width at half maximum of the focused pump spot. We assume that with each absorbed pump photon, one electron-hole pair is injected. The on-axis carrier density excited near the sample surface can then be written as
\begin{equation}
N_0 = \frac{(1-R_0) \alpha_0 F_0}{\hbar \omega},
\end{equation} 
where $\alpha_0$ and $\hbar \omega$ are the absorption coefficient and energy of the pump photon, respectively. The lateral distribution of carriers is Gaussian, identical to the intensity profile of pump spot. From this procedure, we can convert the pump power to injected peak carrier density, by using the complex index of refraction of $5.4 + 3.1j$.\cite{apl96213116}

\begin{figure}
 \includegraphics[width=8.5cm]{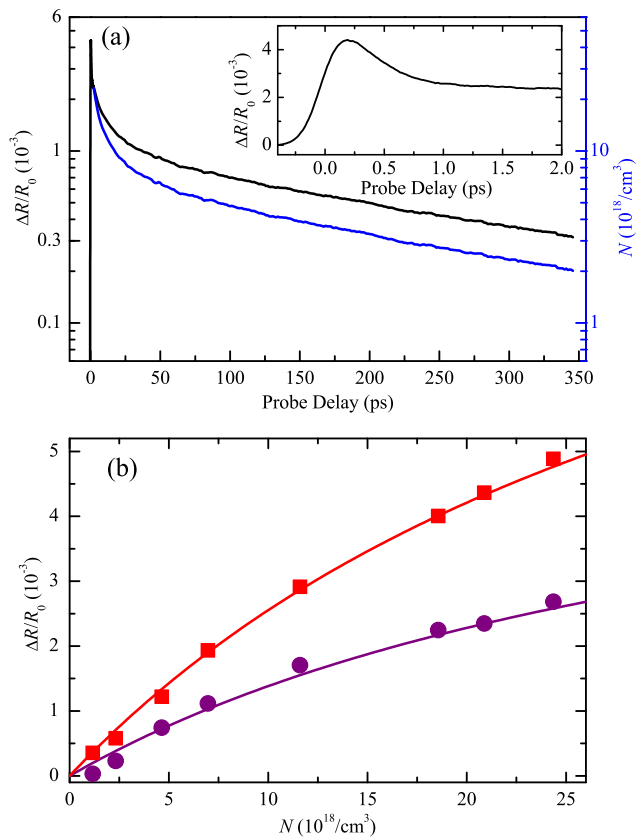}
 \caption{(a) Differential reflection as a function of the probe delay (black curve and left axis). The inset shows the signal in early probe delays. The blue curve (right axis) is the carrier density deduced from the measured differential reflection. (b) Differential reflection as a function of injected carrier density for fixed probe delays of 0.2 (squares) and 2.0 ps (circles). The solid lines are calculated from Eq.~\ref{DRoR} with the same value of $N_s = 3.7 \times 10^{19}$~/cm$^3$.}
  \label{decay}
\end{figure}

\section{Results and discussion}
\label{discussion}

The black curve in Fig.~\ref{decay}(a) (left axis) shows the measured differential reflection signal as a function of the probe delay, with a peak carrier density of $2.1 \times 10^{19}$~/cm$^3$. The inset shows the signal near zero probe delay. Clearly, the dynamics has a fast component of less than 1 ps and a slow one persists for several hundred ps. We attribute the slow process to recombination of carriers, {\it i.e.} carrier lifetime, and the fast process to the dynamical variation of $\sigma$ associated with evolution of carrier distribution in the conduction and valence bands. According to the bandstructure of MoS$_2$, the pump pulse excites electrons from the valance bands to the {\bf K} valley of the conduction band. Since the direct bandgap at {\bf K} point is about 1.84 eV, electrons are excited with an excess energy of about 100 - 200~meV. The pulse-width limited raising time of the differential reflection signal shows that these excited electrons change the excitonic transition strength immediately. This is an evidence that exciton formation is negligible, since one would expect a relatively slow rise of signal associated with exciton formation. Since the {\bf Q} valley in the conduction band is much lower than the {\bf K} valley, these electrons transfer rapidly to the {\bf Q} valley by emitting optical phonons. We attribute the initial decay of the signal in the first ps, with a time constant of about 0.35 ps, to this intervalley transfer process. This intervalley transfer time is comparable to electron phonon scattering time of other types of semiconductors (e.g. GaAs).\cite{l542151} The {\bf Q}-valley electrons still influences the exciton transition, as evidenced by the long-lived signal over 100 ps, but with a smaller cross section. Hence, we attribute the slow dynamics to recombination of carriers in the {\bf Q} valley.

Since the carrier lifetime is on the order of 100 ps, we can assume that carrier density during the first a few ps is a constant, and equal to the injected density. We repeat the measurement shown in the inset of Fig.~\ref{decay} with various pump powers (and hence various injected peak carrier densities). The circles and squares in Fig.~\ref{decay} show the differential reflection signal as a function of the injected carrier density measured at probe delays of 0.2 (the peak) and 2.0 ps, respectively. The dependence is nonlinear, since the injected carrier density is rather high. Both data sets can be fit by using Eq.~\ref{DRoR}, with the same value of $N_s$. By fitting many data sets with various probe delays, we deduce a value of $N_s = (3.7 \pm 0.1) \times 10^{19}$~/cm$^3$, which is used to plot the two solid curves in Fig.~\ref{decay}(b). With the known $N_s$, we can deduce the carrier density as a function of probe delay from the measured differential reflection. The result is plotted as the blue curve in Fig.~\ref{decay}(a) (right axis). Hence, although the differential reflection is a nonlinear function of the carrier density, we can reliably deduce the latter from the former.

Next, we spatially and temporally resolve the differential reflection signal in order to study spatiotemporal dynamics of carriers. The carriers are injected by a focused pump pulse with a Gaussian shape. Hence, the initial carrier distribution is
\begin{equation}
N(r,0)=N_0 \mathrm{exp}[-\frac{r^2}{\sigma^2_0}],
\end{equation}
where $\sigma_0$ is the $1/e$ width of the pump spot. It is related to the full-width at half maxima, $w_0$, by $w_0 = 2 \sqrt{\mathrm{ln}2}\sigma_0$. After injected, carriers diffusion in the sample plane and recombine. Although exciton formation is unlikely, due to the Coulomb attraction, the electron-hole pair still move as a unit in this ambipolar transport process. The dynamics is described by the ambipolar diffusion equation,\cite{b385788}
\begin{equation}
\frac{\partial N(r,t)}{\partial t} = D \nabla^2 N(r,t) - \frac{N(r,t)}{\tau},
\end{equation}
where $D$ and $\tau$ are ambipolar diffusion coefficient and carrier lifetime, respectively. Solving this equation, we have,
\begin{equation}
N(r,t) = \frac{\sigma^2_0 N_0}{\sigma^2 (t) } \mathrm{exp}[\frac{-t}{\tau}] \mathrm{exp} [- \frac{r^2}{\sigma^2 (t) }],
\end{equation}
where $\sigma^2 (t) = \sigma^2_0 + 4 D t$. Hence, we expect the profile remains Gaussian, with the squared width increase linearly with time.

\begin{figure}
 \includegraphics[width=8.5cm]{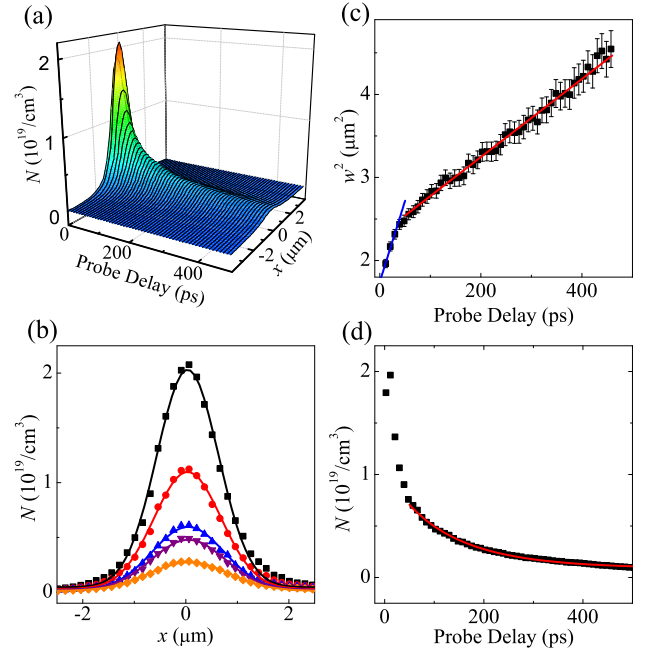}
 \caption{(a) Carrier density as a function of probe delay and $x$ over a long delay range of 500 ps. (b) A few examples of the spatial profiles. The probe delays are (from top to bottom) 10, 30, 75, 100, and 200 ps, respectively. (c) Squared width of the spatial profile of the signal as a function of probe delay. The red line is a linear fit to the data points after 50 ps, which corresponds to a diffusion coefficient of 4.2 cm$^2$/s. The blue line indicates the initial slope, which gives a diffusion coefficient of 18 cm$^2$/s. (d) the carrier density measured at $x=0$ with the finite probe spot. The red line indicates a fit that gives a lifetime of 180 ps.}
  \label{slowdiffusion}
\end{figure}

In order to quantitatively study the carrier dynamics, we measure the Gaussian spatial profiles along $\hat{x}$ axis for many probe delays, with a pump fluence of 10~$\mu$J/cm$^2$. The results are plotted in Fig.~\ref{slowdiffusion}(a). Clearly, the profile becomes lower and wider with increasing the probe delay. Figure~\ref{slowdiffusion}(b) shows a few examples of the measured profiles, with the corresponding Gaussian fits (solid lines). The squared widths deduced from the fits to all the measured profiles are plotted in Fig.~\ref{slowdiffusion}(c). After about 50 ps, the increase becomes linear, as expected from the ambipolar diffusion model. From a linear fit (red line), we deduce a diffusion coefficient of 4.2 $\pm$ 0.4 cm$^2$/s. We note that although the penetration depth is only on the order of 20 nm due to strong absorption, the diffusion of carriers along perpendicular direction is expected to be much slower due to the layered structure of MoS$_2$.

We are not aware of any previous studies on charge carrier diffusion in bulk MoS$_2$. However, from the measured ambipolar diffusion coefficient, we can estimate the carrier mobility and compare with previous measurements on mobility. Based on the fact that the effective masses of electrons and holes in MoS$_2$ are similar,\cite{ieeeted583042}
we assuming that the mobilities of electrons and holes are comparable. Therefore, the ambipolar diffusion coefficient can be approximately treated as unipolar diffusion coefficients. Using the Einstein relation, $D/k_BT = \mu/e$, where $k_B$, $T$, $\mu$, and $e$ are the Boltzmann constant, the temperature, the mobility, and the absolute value of the electron charge, the measured $D$ after 50 ps corresponds to a mobility of 170 $\pm$ 20~cm$^2$/Vs. Early electrical measurements have deduced mobilities in the range of 100 - 260~cm$^2$/Vs in nominally undoped bulk MoS$_2$ crystals with carrier densities in the range of $10^{15}$ to $10^{16}$ /cm$^3$.\cite{pr163743} Our result is reasonably consistent with these measurements. Furthermore, in the first 50 ps, the squared width increases sub-linearly, with a decreasing slope. The initial slope, indicated by the blue line in Fig.~\ref{slowdiffusion}(c), corresponds to a diffusion coefficient of 18 cm$^2$/s. We attribute this to hot carrier effect. If the mobilities is not a strong function of temperature, hot carriers with larger thermal velocities diffuse faster. Hence, the sub-linear increase of the squared width in the first 50 ps reflects the cooling of carriers, and provide a direct evidence of energy relaxation of hot carriers. Such a hot carrier diffusion has been observed in other types of semiconductors, such as ZnSe.\cite{l94137402,b67035306}

The spatiotemporal resolution of the carrier density also allow us to deduce the carrier lifetime. Figure~\ref{slowdiffusion}(d) shows the measured carrier density as a function of probe delay with the probe spot overlapped with the pump spot ($x = 0$). The decrease of the carrier density is caused by both the carrier recombination and the transport of carriers out of the probing area. To deduce the carrier lifetime, we consider that the intensity of the Gaussian-shaped probe spot, with the center overlapped with the center of the carrier density profile, is
\begin{equation}
I(r)=I_0 \mathrm{exp}[-\frac{r^2}{\sigma^2_p}],
\end{equation}
where $I_0$ is on-axis peak intensity, and $\sigma_p$ is the $1/e$ width of the probe spot. By integrating the product of such a sensitivity function with the carrier density profile, we find that the carrier density measured by such a probe,
\begin{equation}
N(x=0) \propto \frac{1}{\sigma^2_p + \sigma^2_0 + 4 D t} \mathrm{exp}[\frac{-t}{\tau}],
\label{decayx0}
\end{equation}
Here, the first factor describes the effect of diffusion of carriers out of probing area, and the second factor accounts for recombination. We use Eq.~\ref{decayx0} to fit the data shown in Fig.~\ref{slowdiffusion}(d), with the known value of $D$. We obtain a satisfactory agreement with the data (red curve), and deduce a carrier lifetime of $180 \pm 20$~ps. We note that since MoS$_2$ has an indirect bandgap, the radiative recombination of carriers is restricted by crystal momentum conservation. Accordingly, one would expect a long carrier lifetime due to radiative recombination. The short lifetime deduced here is probably limited by nonradiative recombination at defects and surface. It is also interesting to note that such a lifetime is consistent with that in atomically thin film MoS$_2$ measured by time-resolved photoluminescence.\cite{apl99102109}

\section{Summary}

We have used a transient absorption microscope to study charge carrier dynamics in bulk MoS$_2$ crystals, and reveals several aspects of the charge carrier dynamics. We find that after the carriers are injected to the more energetic {\bf K} valley, they transfer to the low energy {\bf Q} valley with a time constance of about 0.35 ps. After that, the carriers relax their energy in about 50 ps until they reach thermal equilibrium with the lattice. The rest of the carrier dynamics can be well described by ambipolar diffusion of thermalized carriers, with a diffusion coefficient of 4.2~$\pm$~0.4~cm$^2$/s and a lifetime of $180 \pm 20$~ps. These results provide fundamental information on charge carrier dynamics in bulk MoS$_2$ crystals, and can be used to understand carrier dynamics of MoS$_2$ bulk and atomic layers.

\section{Acknowledgement}
We acknowledge support from the US National Science Foundation under Awards No. DMR-0954486 and No. EPS-0903806, and matching support from the State of Kansas through Kansas Technology Enterprise Corporation. We also acknowledge support from the National Basic Research Program 973 of China: 2011CB932700, 2011CB932703; Chinese Natural Science Fund Project under Contract No. 61077044; and Beijing Natural Science Fund Project under Contract No. 4132031.

%\bibliography{/users/huizhao/Documents/Bibfile/literature.bib}

%merlin.mbs aipnum4-1.bst 2010-07-25 4.21a (PWD, AO, DPC) hacked
%Control: key (0)
%Control: author (8) initials jnrlst
%Control: editor formatted (1) identically to author
%Control: production of article title (-1) disabled
%Control: page (0) single
%Control: year (1) truncated
%Control: production of eprint (0) enabled
%

\end{document}